\documentclass[nonacm,sigconf,screen,10pt]{acmart}
\usepackage{graphicx}
\usepackage{hyperref}
\usepackage[frozencache,cachedir=minted-cache]{minted}
\usepackage{xcolor}
\usepackage{tabularray}
\usepackage{xspace}
\usepackage{subcaption}
\usepackage{enumitem}

\newenvironment{mylisting}[1]{%
    \VerbatimEnvironment
    \begin{minted}[
        frame=single,
        framesep=3pt,
        fontsize=\footnotesize,
        fontseries=b,
        breaklines=true,
        linenos=true,
        numbersep=3pt,
        style=xcode
    ]{#1}%
}{%
    \end{minted}%
}

\def\Snospace~{\S{}}

\newcommand{\rattan}{\textsc{Rattan}\xspace}

\title{\rattan: An Extensible and Scalable Modular Internet Path Emulator}

\author{Minhu Wang}
\authornote{Equal contributions}
\affiliation{
    \institution{Tsinghua University}
    \country{}
}

\author{Yixin Shen}
\authornotemark[1]
\affiliation{
    \institution{Tsinghua University}
    \country{} 
}

\author{Bo Wang}
\affiliation{
    \institution{Tsinghua University}
    \country{} 
}

\author{Haixuan Tong}
\affiliation{
    \institution{Tsinghua University}
    \country{} 
}

\author{Yutong Xie}
\affiliation{
    \institution{Tsinghua University}
    \country{} 
}

\author{Yixuan Gao}
\affiliation{
    \institution{Tsinghua University}
    \country{} 
}

\author{Yan Liu}
\affiliation{
    \institution{ByteDance}
    \country{}
}

\author{Li Chen}
\affiliation{
    \institution{ByteDance}
    \country{}
}

\author{Mingwei Xu}
\affiliation{
    \institution{Tsinghua University}
    \country{} 
}

\author{Jianping Wu}
\affiliation{
    \institution{Tsinghua University}
    \country{} 
}

\begin{abstract}
The rapid growth of Internet paths in heterogeneity, scale, and dynamics has made existing emulators increasingly insufficient in flexibility, scalability, and usability. 
To address these limitations, we present \rattan{}, an extensible and scalable software network path emulator for modern Internet conditions. 
\rattan{}'s core innovation lies in its cell-based architecture: by splitting emulation functions into modular ``cells'' with well-documented asynchronous interfaces, users are allowed to easily compose different cells by hierarchically linking them and easily construct new cells by using standard cell interfaces. 
This design enables: 
(1) scalability, supporting hundreds of concurrent gigabit-level paths on a single machine and cluster-level experiments composed of multiple machines; 
(2) extensibility, simulating new network conditions by constructing new cells. 
\rattan{} empowers developers and researchers to efficiently and confidently evaluate, validate, and diagnose diverse network transport innovations for online services.
\end{abstract}

\begin{document}
\maketitle

\section{Introduction}
Modern large-scale online services operate over billions of heterogeneous and dynamic \textit{network paths}: the communication paths between billions of users and online services span diverse infrastructures, protocols, and policies.
At this large scale, even minor service degradation affecting a small subset of users can cause significant revenue loss\ \cite{MillisecondsMakeMillions}. This reality imposes strict requirements on the Quality of Experience (QoE) of these services and, in turn, on the underlying network Quality of Service (QoS).
Meanwhile, these same economies of scale also amplify the potential benefits of QoS improvements tailored to specific network conditions. This prospect has thus spurred a wave of innovations across the transport stack, including new transport protocols\ \cite{DBLP:conf/sigcomm/LangleyRWVKZYKS17}, multi-path scheduling\ \cite{rfc8684, DBLP:conf/sigcomm/ZhengMLYLZZSCLA21}, advanced congestion control\ \cite{DBLP:journals/queue/CardwellCGYJ16,DBLP:conf/nsdi/ArunB18,DBLP:conf/sigcomm/YenAC23}, and adaptive bitrate selection\ \cite{DBLP:conf/mmsys/Stockhammer11,DBLP:conf/sigcomm/MaoNA17}.

However, a fundamental tension arises between the drive to rapidly deploy these innovations and the need to maintain robust network performance: we must ensure these innovations deliver real benefits in their target scenarios while preventing performance degradation in all other cases. This tension necessitates an efficient, cost-effective, and reliable tool to evaluate, validate, and diagnose the performance of network transport innovations before deploying them to the highly diverse Internet.

Emulation~\cite{DBLP:conf/nsdi/SanagaDRL09, DBLP:conf/hotnets/LantzHM10} has long been a popular choice for this purpose. It creates controllable, artificial network environments that selectively reproduce transport characteristics (e.g., bandwidth, delay, reordering, and random drops) of actual Internet paths~\cite{DBLP:conf/tridentcom/PizzoniaR08}. By strategically selecting emulation models that focus on key characteristics while ignoring irrelevant details, emulation delivers high computational efficiency without sacrificing accuracy. The ability to iterate through arbitrary network conditions overcomes the cost and potential sampling bias of small-scale, real-world testbeds~\cite{DBLP:conf/imc/SpangHKHMJ21}. Furthermore, unlike simulation, emulation allows for the direct testing of unmodified services with their production traffic, rather than modeling and reimplementing them in a simulation framework~\cite{DBLP:books/sp/wehrle2010/RileyH10}. This approach eliminates potential modeling errors and reduces development overhead.

Despite these advantages, it is increasingly difficult for emulators to accurately and efficiently mimic the transport properties of the rapidly evolving Internet. As Internet paths become faster, more heterogeneous, and more dynamic (\autoref{sec:background-evolving-internet-paths}), it is increasingly challenging for network developers to create new emulation models, optimize emulation to run at line rate, and conduct large-scale experiments covering all desired network conditions using existing emulators (\autoref{sec:background-whats-wrong-with-existing-emulators}).

To address these challenges, we present \rattan{}, a high-performance modular framework for large-scale network path emulation targeting end-host services, with the following characteristics:

\noindent\textbf{High Performance.} \rattan{} achieves exceptional emulation performance through its asynchronous architecture with optimized packet I/O and efficient multi-core scheduling. It supports the emulation of paths serving flows with speeds reaching tens of gigabits per second.

\noindent\textbf{Flexibility.} \rattan{}'s core logic is designed to be agnostic to the underlying path emulation model. Emulation models are implemented in basic modules called ``cells'' with well-documented asynchronous interfaces. By hierarchically linking cells, \rattan{} allows for the flexible combination of arbitrary emulation models, from the simplest queue-based link emulation to sophisticated learned neural networks\ \cite{DBLP:conf/hotnets/AshokDNPSG20,DBLP:journals/pomacs/AshokTNPS22,DBLP:conf/aaai/AnshumaanBTNSP23}.

\noindent\textbf{Extensibility.} \rattan{} users can construct new network conditions by combining cells that cover widely-used emulation models provided by \rattan{}. They can also easily write new cells to add custom emulation models using standard cell interfaces. User-defined cells get full support in \rattan{} and can be used interchangeably with built-in cells.

\noindent\textbf{Scalability.} Due to its efficient architecture, \rattan{} supports dense deployments of thousands of concurrent emulated megabit-level paths on a single machine, facilitated by a per-machine central resource manager. Furthermore, \rattan{} implements an orchestrator for cluster-level task scheduling to achieve even higher levels of parallelism.

\noindent\textbf{User Friendliness.} \rattan{} provides both a library supporting finer-granularity control and a command-line tool optimized for common use cases. Users can provision common emulation environments with a single command or customize every detail through the library.

In summary, \rattan{} empowers developers and researchers to efficiently and confidently evaluate, validate, and diagnose network performance across the full spectrum of modern Internet conditions. In the remainder of this paper, we further discuss the background of Internet evolution and the limitations of existing emulators in \autoref{sec:background}, describe \rattan{}'s design in more detail in \autoref{sec:rattan-design}, and demonstrate its effectiveness with a case study in \autoref{sec:rattan-case-study}.

\section{Background \& Motivation}\label{sec:background}
\subsection{The Evolving Internet}\label{sec:background-evolving-internet-paths}
Also driven by the demands of large-scale online services, the Internet continues to evolve at an unprecedented pace.

First, the Internet is becoming \textit{faster}. Decades of investment in network infrastructure and advancements in access technologies have significantly increased Internet access speeds worldwide. As Ookla's open database\ \cite{ookla_speedtest_2024} shows, from 2019 to 2024, the global median mobile download speed increased from $16.5\ \text{Mbps}$ to $55.9\ \text{Mbps}$, while the maximum speed increased from $804.1\ \text{Mbps}$ to $4797.9\ \text{Mbps}$. Since emulators must intercept, process, and forward every packet from unmodified applications in real-time while applying emulation effects, these high-speed paths demand emulators with proportionally \textbf{high performance}. In extreme cases, emulators must be capable of handling traffic up to tens of gigabits per second to faithfully represent network segments serving multiple competing flows at cutting-edge speeds.

Furthermore, the Internet is becoming increasingly \textit{heterogeneous} as every network layer grows more diverse. At the physical layer, there are more access technologies and transmission media (e.g., 5G~\cite{andrews2014will}, satellites~\cite{michel2022first}, and power lines~\cite{lin2002power}) with distinct physical properties and medium access control mechanisms. For example, cellular networks have been found to allocate more bandwidth to ``heavy'' workloads with longer and larger flows\ \cite{DBLP:conf/nsdi/Sentosa0GH25}. At the network layer, operators employ richer routing and traffic management policies to enforce their resource management strategies~\cite{DBLP:conf/www/ZhangJHHM0Z24, DBLP:conf/sigcomm/LiNCGM19, DBLP:conf/imc/WangLS24}. For example, mobile ISPs may install explicit rate limiters that severely restrict the maximum bandwidth for flows from certain users, leading to severe packet drops\ \cite{DBLP:conf/icnp/ZhuLGNL23, DBLP:conf/sigcomm/LiNCGM19}. At the transport and application layer, flows' reactions to network congestion signals, jointly decided by traffic patterns of applications and rate control behaviors of transport protocols \cite{DBLP:conf/sigcomm/Goyal0CNAB22, DBLP:conf/sigcomm/MengSGS20}, also complicate Internet paths. For example, the bandwidth one flow can acquire from a full link depends on the aggressiveness of congestion control algorithms (CCAs) employed by competing flows sharing the path. ~\cite{DBLP:conf/hotnets/ZapletalK23, DBLP:conf/imc/WareMSS19, DBLP:conf/hotnets/BrownK0KPSS23}.

The transport performance effects of these diverse factors and their interactions create a complex emulation landscape far beyond what traditional tools were designed to handle. This multifaceted complexity challenges emulators in two ways. First, emulators must be \textbf{flexible} enough to support and combine domain-specific emulation models to provide meaningful results. Second, they must be sufficiently \textbf{scalable} to support large sets of parallel experiments for comprehensive coverage of the potential network conditions of online services within a reasonable time.

What's more, the Internet is becoming increasingly \textit{dynamic}. The growing popularity of Software-Defined Networking (SDN) makes it more convenient for operators to apply route changes and traffic engineering policy to their networks \cite{DBLP:conf/sigcomm/JainKMOPSVWZZZHSV13}. Occasional errors across the network may also cause unexpected changes to path properties. Evidence in large-scale measurement \cite{DBLP:conf/usenix/YanMHRWLW18} and production environment shows that path properties may obviously change in days to weeks~\cite{DBLP:conf/nsdi/ZengXCZ00CQZHZ25}. To emulate these fast-changing network conditions, emulators must be highly \textbf{extensible} to allow for the easy incorporation of new emulation models and \textbf{user-friendly} so that users can spend more time on performance analysis instead of implementing new emulators.

\begin{figure*}[ht]
    \includegraphics[width=\textwidth]{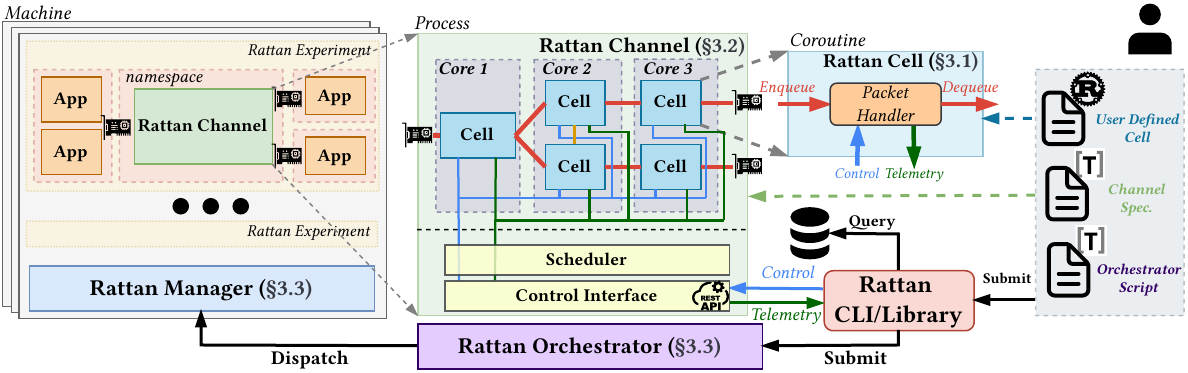}
    \caption{Overview of \rattan{}}\label{fig:rattan-design}
\end{figure*}

\subsection{Limitations of Existing Emulators}\label{sec:background-whats-wrong-with-existing-emulators}
Unfortunately, existing emulators fail to meet the demands for high performance, flexibility, scalability, extensibility, and user-friendliness that are essential to support the development of modern online services.

The most conventional link emulators embedded in operating system kernels (e.g., DummyNet\ \cite{DBLP:journals/ccr/Rizzo97} for BSD and NetEm\ \cite{hemminger2005network} for Linux) and the network emulation tools built on top of them (e.g., ModelNet\ \cite{DBLP:conf/osdi/VahdatYWMKCB02}, Emulab\ \cite{DBLP:conf/usenix/HiblerRSDGSWL08}, and Mininet\ \cite{DBLP:conf/hotnets/LantzHM10}) are limited in performance due to high kernel packet processing overheads\ \cite{DBLP:conf/conext/Hoiland-Jorgensen18}. They are also inflexible due to very limited emulation model support, and not extensible due to the difficulty of kernel programming. While most of them are easy to use on a single machine, they lack support for larger-scale deployments.

Later userspace software emulators, such as MahiMahi\ \cite{DBLP:conf/usenix/NetravaliSDGWMB15} and Pantheon\ \cite{DBLP:conf/usenix/YanMHRWLW18} (built on MahiMahi), implement more emulation models (e.g., link trace recording and replay). Extending their functionality is feasible but still requires significant hand-coding due to the lack of an extensible architecture and a well-documented interface. Recent research has also presented several domain-specific emulation models and prototypes, such as iBox\ \cite{DBLP:conf/hotnets/AshokDNPSG20,DBLP:journals/pomacs/AshokTNPS22,DBLP:conf/aaai/AnshumaanBTNSP23}. These tools are direct responses to the ever-increasing complexity of network conditions. However, their prototypes are usually tightly coupled to their models and therefore lack extensibility.

Hardware network emulators generally offer excellent (usually line-rate) performance with the help of highly customized hardware (e.g., ASICs or FPGAs), which nonetheless limits their extensibility and scalability. Implementing new emulation models on their closed hardware systems while maintaining performance advantages is difficult or even impossible. Meanwhile, high equipment costs restrict their scalability.

The significant gap between the emulation capabilities that network developers require and what existing tools can provide motivated us to design a new emulator from scratch.

\section{\rattan{} Design}\label{sec:rattan-design}

In this section, we introduce \rattan's hierarchical and modular design shown in \autoref{fig:rattan-design} in a bottom-up manner, from its basic building block (cell, \autoref{sec:rattan-cell}), minimal functional unit (channel, \autoref{sec:rattan-channel}), higher-level task manager (manager and orchestrator, \autoref{sec:rattan-manager-orchestrator}) to show how it achieves extensibility and scalability.

\subsection{\rattan{} Cell}\label{sec:rattan-cell}
\begin{figure*}[ht]
    \begin{mylisting}{rust}
struct DelayCell { internal_queue: VecDeque<P>, timer: Timer, config: DelayCellConfig, stats: DelayCellStats }
impl RattanCell for DelayCell {
    fn enqueue(&self, mut packet: P) { self.internal_queue.push_back(packet) }
    async fn dequeue(&self) -> Option<P> {
        let packet = self.internal_queue.pop_front()?;
        self.timer.sleep(packet.get_timestamp() + self.delay - Instant::now()).await;
        self.stats.update(&packet);
        Some(packet) 
    }
    async fn control(&self, config: Self::Config) { self.config.update(config).await; }
    async fn telemetry(&self) -> DelayCellStats { self.stats.collect().await }
}
    \end{mylisting}
    \caption{An example of a \rattan cell that adds a configurable propagation delay to all passing packets.}\label{fig:rattan-cell-example}
\end{figure*}

In \rattan{}, complex network conditions are emulated by hierarchically composing small building blocks called \textit{cells}. Each cell applies a unique, custom effect on forwarded packets to emulate certain properties of a network path. 

Possible functionalities of cells could be simple features available in existing emulators, such as throttling forwarding speed based on a recorded link trace to emulate bottleneck link capacity, delaying packets by a constant time to emulate the stable propagation delay of long-distance fiber links, and periodically dropping packets to emulate loss patterns on wireless links. Furthermore, Cells can emulate more complex behaviors, incorporating deeper domain knowledge, such as adding variable delays that model WiFi-layer retransmissions, imposing dynamic capacity and delay according to real-time traffic patterns \cite{DBLP:conf/nsdi/Sentosa0GH25}, or using a physics-based model to emulate the link dynamics induced by the motion of Low-Earth Orbit (LEO) satellites.

Cells implement their functionalities following the \textit{actor concurrency model} \cite{DBLP:phd/us/Agha85} for better scalability and efficiency. Specifically, each cell runs as a self-contained actor, sharing no state with other cells and interacting with the outside world through four standard message-passing interfaces: \textbf{enqueue}, which receives packets from other cells; \textbf{dequeue}, which sends processed packets to other cells at a time determined by the cell; \textbf{control}, which receives updated configurations and other custom commands from the user at runtime; and \textbf{telemetry}, which exposes custom internal states (e.g., traffic statistics, packet traces, and emulation logs) to the user. We show an example of a cell implemented in the Rust programming language\footnote{We develop \rattan{} with Rust for its strong type system and excellent asynchronous programming support. Currently, Rust is the only supported language for implementing custom cells. We plan to provide cell interface libraries for other languages in the future.} in \autoref{fig:rattan-cell-example}.

As shown in \autoref{fig:rattan-cell-example}, \rattan{} employs an asynchronous execution model for cells, simplifying their implementation by eliminating the need for manual synchronization when handling events from various sources (e.g., packets, configuration updates, and telemetry requests). This model also facilitates more efficient scaling and scheduling of cells (discussed in \autoref{sec:rattan-channel}).

Although we use a conventional emulator implementation with an internal queue and customized enqueue/dequeue policies in \autoref{fig:rattan-cell-example}, \rattan{} imposes no restrictions on cell implementations other than the basic asynchronous rule that a cell must not block its execution thread. Cells can freely implement their desired emulation models. For instance, for a learned reactive model, cells can create dedicated computation threads for deep generative models that synthesize high-quality network traces in real time \cite{DBLP:conf/aaai/AnshumaanBTNSP23}. They can even communicate with other processes or remote services to incorporate proprietary middleboxes.

\rattan{} provides an extensive library of widely used \textit{cells} to simplify the construction of common emulation scenarios. For a unified architecture, \rattan{} also implements its internal functionalities as cells. For example, NIC cells are responsible for receiving and transmitting packets on physical or virtual NICs, while router cells distribute traffic to multiple downstream cells according to configurable rules. Built-in, special-purpose, and user-supplied cells are all treated equally in \rattan{}.

\subsection{\rattan{} Channel}\label{sec:rattan-channel}
A channel is the minimal functional and resource management unit in \rattan{}. It employs a control- and data-plane separation design, with a data plane consisting of a directed acyclic graph (DAG) of cells (\autoref{sec:rattan-cell-data-plane}) and a control plane, called the channel runtime, for managing the execution of these cells (\autoref{sec:rattan-channel-control-plane}).

\subsubsection{Data Plane}\label{sec:rattan-cell-data-plane} 
\begin{figure}[ht]
    \begin{subfigure}[b]{\linewidth}
        \includegraphics[width=\textwidth]{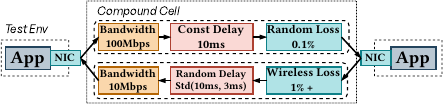}
        \caption{\raggedright An emulated path with asymmetric up and down links.}\label{fig:rattan-example-1}
    \end{subfigure}
    \begin{subfigure}[b]{\linewidth}
        \includegraphics[width=\textwidth]{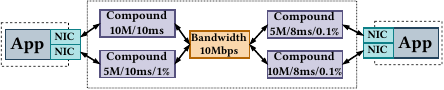}
        \caption{\raggedright An emulated multi-path condition with a shared bottleneck.}\label{fig:rattan-example-2}
    \end{subfigure}
    \begin{subfigure}[b]{\linewidth}
        \includegraphics[width=\textwidth]{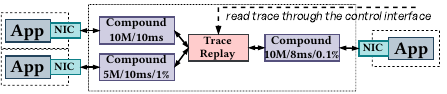}
        \caption{\raggedright An emulated path combining link emulation and trace replay.}\label{fig:rattan-example-3}
    \end{subfigure}
    \caption{Examples of \rattan{} channels demonstrating how cells can be combined to emulate advanced network conditions.}\label{fig:rattan-examples}
\end{figure}

As shown in \autoref{fig:rattan-examples}, users can flexibly combine cells with individual functionalities to build the complex network conditions needed by recent network research.

For instance, a fixed broadband Internet path can be modeled by chaining together bandwidth, delay, and loss cells (which we refer to as a compound cell), similar to the approach in existing emulators such as NetEm and MahiMahi. Furthermore, users can build a path with asymmetric uplink and downlink channels using two such compound cells (\autoref{fig:rattan-example-1}), construct scenarios where two seemingly independent paths share a bottleneck link (\autoref{fig:rattan-example-2}), or combine link emulation and trace replay to emulate cases where a Wi-Fi user and a wired user compete for a common residential uplink, as illustrated in \autoref{fig:rattan-example-3}.

\begin{figure*}[ht]
    \begin{mylisting}{rust}
let mut ch_config = RattanConfig::<StdPacket> { env: StdNetEnvConfig { endpoints: 2, symmetric: true } };
ch_config.cells.insert("bw", (BwCellConfig { bw: "100Mbps", queue: QueueConfig { policy: "droptail", len: 1000} }));
ch_config.cells.insert("delay", (DelayCellConfig { delay: "10ms" }));
ch_config.cells.insert("loss", (LossCellConfig { drop_rate: 0.01 }));
ch_config.links = HashMap::from([("end0", "bw"), ("bw", "delay"), ("delay", "loss"), ("loss", "end1")]);
RattanChannel::<AfPacketDriver>::new(ch_config).run();  // run the channel directly
RattanManagerClient::new().connect().submit(ch_config); // or submit to the Rattan manager
    \end{mylisting}
    \caption{An example of a \rattan channel emulating a single link with constant bandwidth, delay, and loss by chaining predefined bandwidth, delay, and loss cells.}\label{fig:rattan-channel-example}
\end{figure*}

\subsubsection{Control Plane}\label{sec:rattan-channel-control-plane} 
The \rattan{} channel's control plane has three responsibilities: exposing the control interface, scheduling cell execution, and detecting emulation errors.

The runtime automatically generates standard REST HTTP endpoints for all cells based on their \texttt{control} and \texttt{telemetry} interfaces. Users can control cell behavior and retrieve internal states through these endpoints during emulation.

Like a conventional asynchronous executor, the runtime maintains a local run queue to sort all cell coroutines by their next scheduled execution time (i.e., packet sending time). It employs a standard I/O multiplexing interface (e.g., \texttt{epoll}) to monitor external events, including incoming NIC packets, HTTP requests containing user commands or telemetry queries, and other asynchronous I/O events registered by cells. For each external event, the runtime calls the corresponding interfaces of the target cells to handle it. The scheduler executes all runnable cells in a round-robin fashion, following their topological order in the DAG.

For low-rate emulation, cells spend most of their time waiting for the next opportunity to send packets, allowing for the simultaneous operation of multiple channels on a single core. For extremely high-speed emulation, however, \rattan{} can scale to multiple cores, potentially assigning a dedicated core to each cell. \rattan{} employs a custom multi-core scheduler to balance emulation accuracy and resource usage. Specifically, the DAG is partitioned into multiple subgraphs and assigned to different physical cores. The local scheduler on each core manages its subgraph and reports cell utilization to a global scheduler, which rebalances workloads across cores by re-partitioning the DAG using the CPU usage of cells as weights.

The runtime monitors the scheduling delay of each cell (i.e., the time gap between when a cell becomes runnable and when it is actually executed) as an indicator of emulation accuracy. It issues a warning to the user when this delay exceeds a configurable threshold. The user can then increase the number of cores or reduce the number of parallel channels to resolve the issue, or at least be aware that the emulation results may not be sufficiently accurate.

\subsubsection{Channel Usage}
As shown in \autoref{fig:rattan-channel-example}, users can write channels interactively using the \texttt{librattan} Rust library. \rattan{} also supports implementing channels declaratively with a configuration file that defines cells and links.

From an outside view, a \rattan{} channel is a user-space process running inside a dedicated network namespace, forwarding packets between an arbitrary number of virtual or physical NICs. Unmodified client and server software (which may also be isolated in their own network namespaces for a clean network environment) can communicate through these NICs to use the emulation environment without any knowledge of \rattan{}'s internals.

One interesting way to utilize \rattan{} is by mixing a real network path with an emulated one. \rattan{} can forward packets between a virtual NIC exposed to an application on one end and a physical NIC to send traffic through the Internet on the other.

The \rattan{} channel is designed to be stateless from the OS perspective and runs on given system resources, such as network namespaces and virtual interfaces. It also never modifies system states, such as the routing tables and NAT rules required for the aforementioned hybrid emulation. State management is delegated to the machine-level \rattan{} manager. This approach simplifies channel implementation and the management of parallel channels sharing the same machine.

\subsection{\rattan{} Manager \& Orchestrator}\label{sec:rattan-manager-orchestrator}
As shown in \autoref{fig:rattan-design}, the \rattan{} manager is an optional component that manages system resources for \rattan{} channels. It sets up resources like network namespaces, virtual interfaces, routing tables, and firewall rules for channels. Additionally, it attempts to reuse these system resources across experiments to minimize provisioning overhead.

On the cluster level, we also design a task planner and dispatcher, called the \rattan{} orchestrator, to help conduct large-scale experiments. Users can launch a batch of experiments with an orchestrator script specifying workloads and emulation channels. The orchestrator will then notify managers (running as a system service) on all worker machines to pull experiments from it. Managers will pull experiments according to their available hardware resources, conduct the experiments with constructed \rattan{} channels, and write results to a database for later analysis.

\begin{figure*}[ht]
    \begin{subfigure}[b]{0.32\textwidth}
        \includegraphics[width=\textwidth]{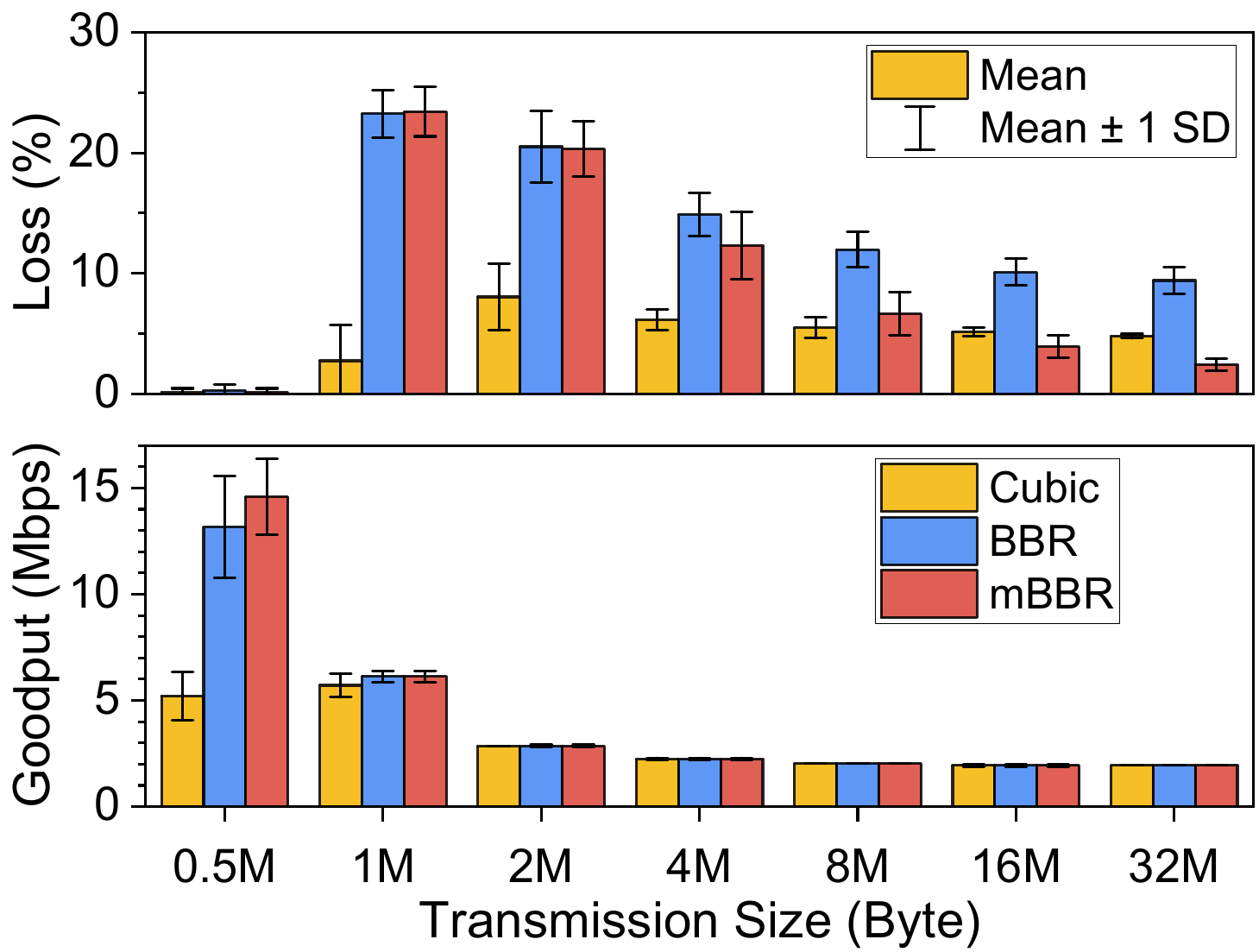}
        \caption{[Real Mobile Network]}
        \label{fig:rattan-case-study-a}
    \end{subfigure}
    \hfill
    \begin{subfigure}[b]{0.32\textwidth}
        \includegraphics[width=\textwidth]{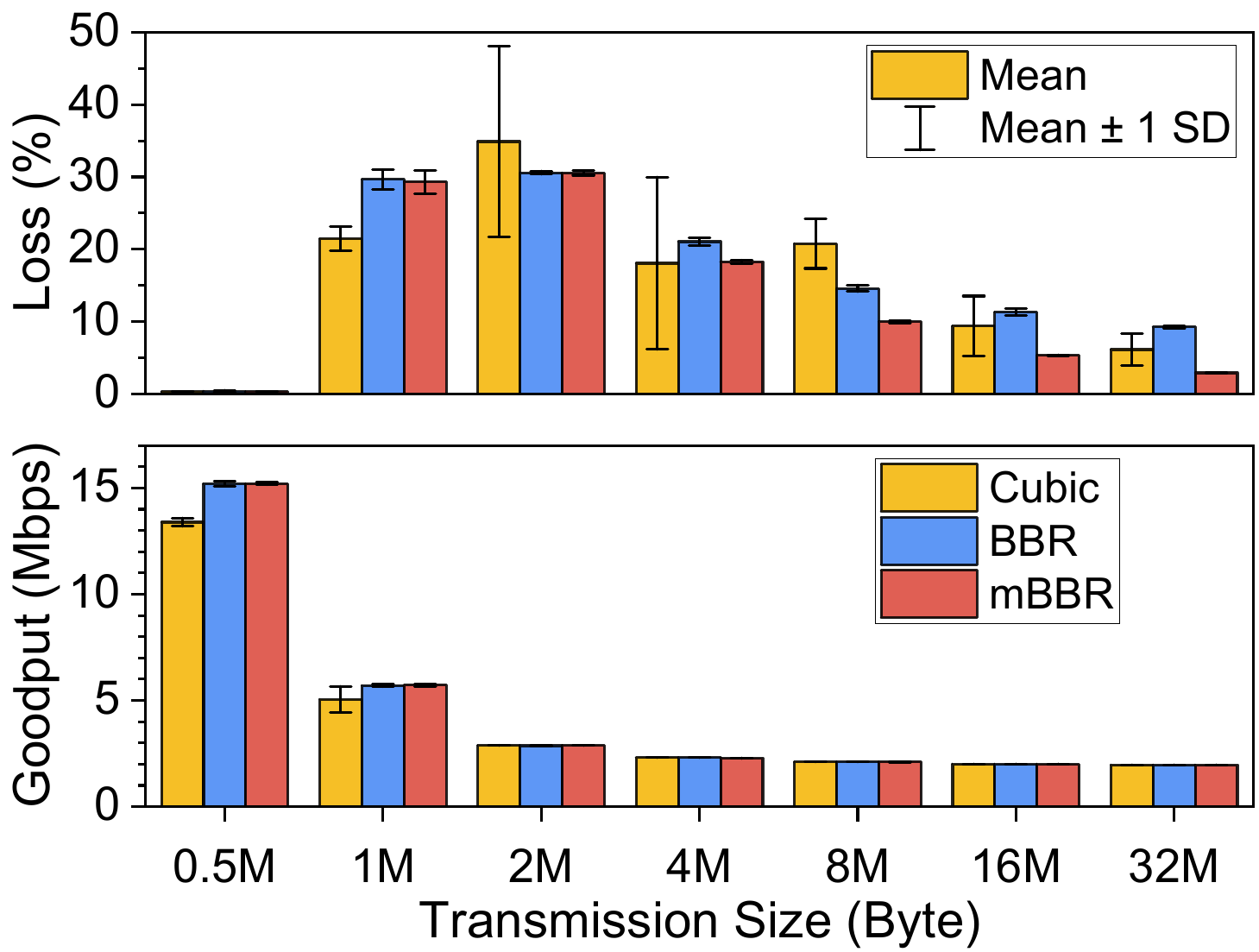}
        \caption{[mBBR Emulation]}
        \label{fig:rattan-case-study-b}
    \end{subfigure}
    \hfill
    \begin{subfigure}[b]{0.32\textwidth}
        \includegraphics[width=\textwidth]{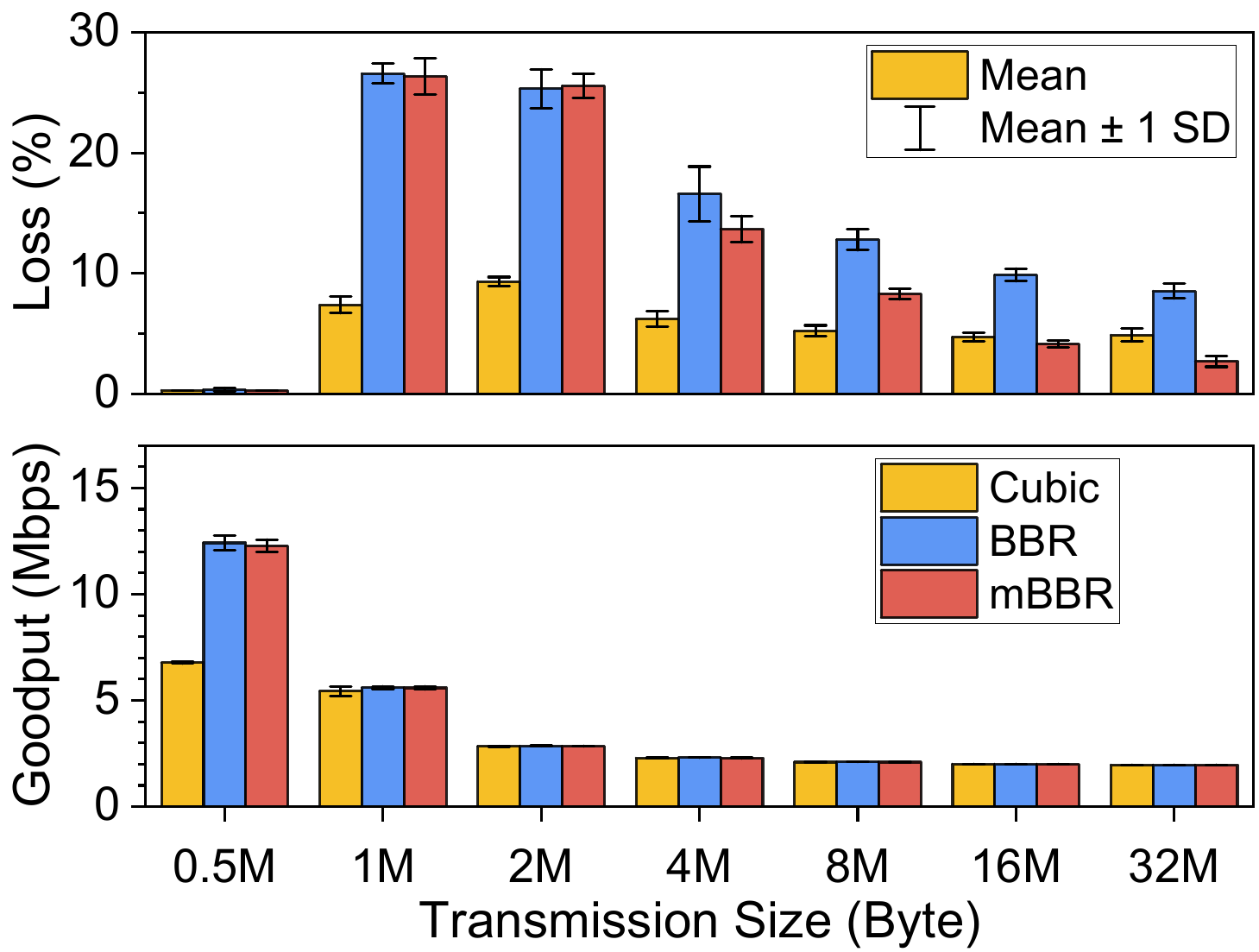}
        \caption{[\rattan{} Emulation]}
        \label{fig:rattan-case-study-c}
    \end{subfigure}
    \caption{A comparison of results reproducing the mBBR experiment in three different environments: a real mobile network, mBBR's emulation approach, and \rattan{}'s emulation approach.}\label{fig:rattan-case-study}
\end{figure*}

\begin{figure}[ht]
    \includegraphics[width=0.65\linewidth]{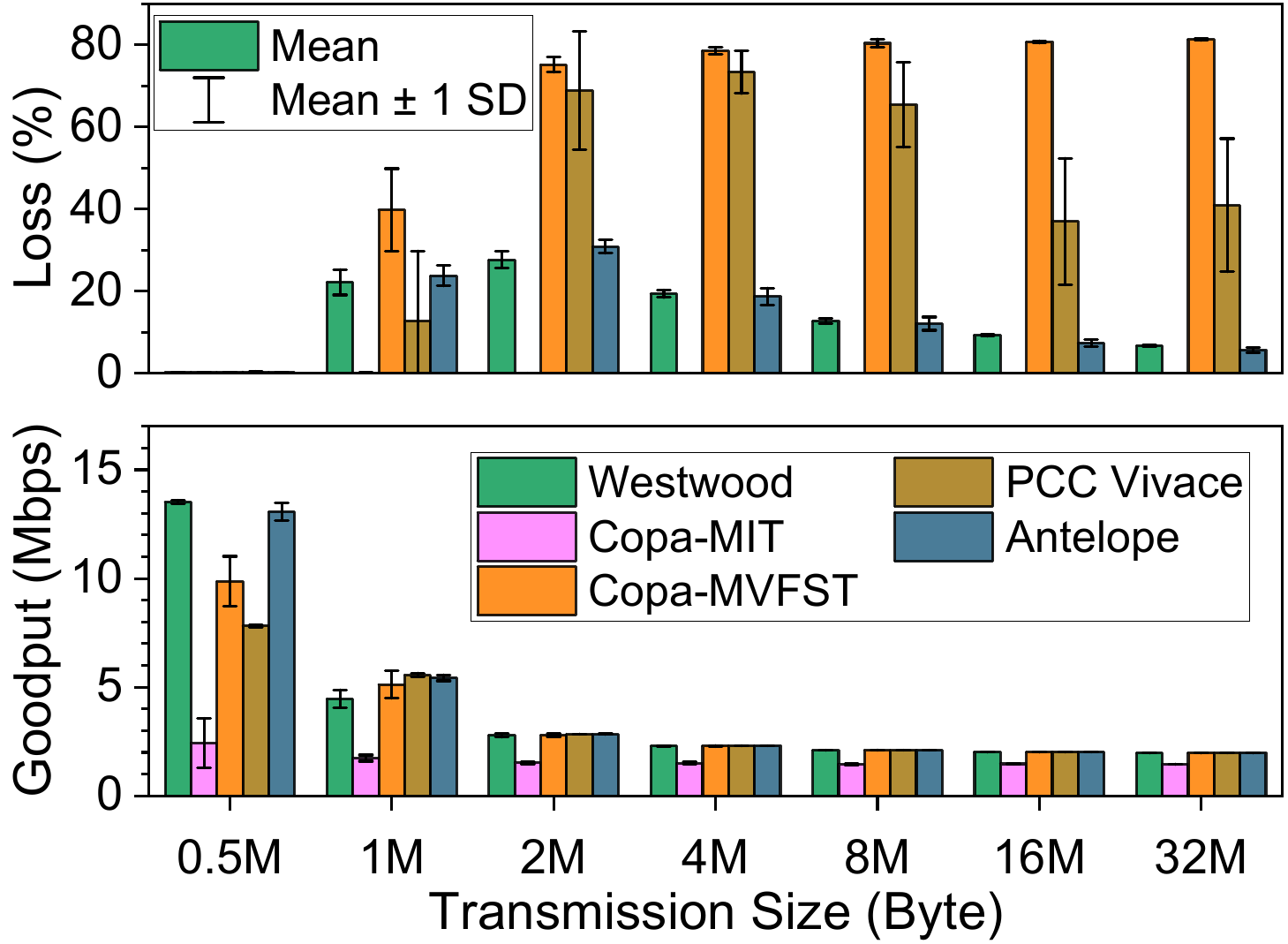}
    \caption{[\rattan{} Emulation] Goodput and packet loss rates of five additional CCAs on a mobile network with traffic policing.}\label{fig:rattan-case-study-different-tokenbucket}
\end{figure}

\begin{figure}[ht]
    \includegraphics[width=0.8\linewidth]{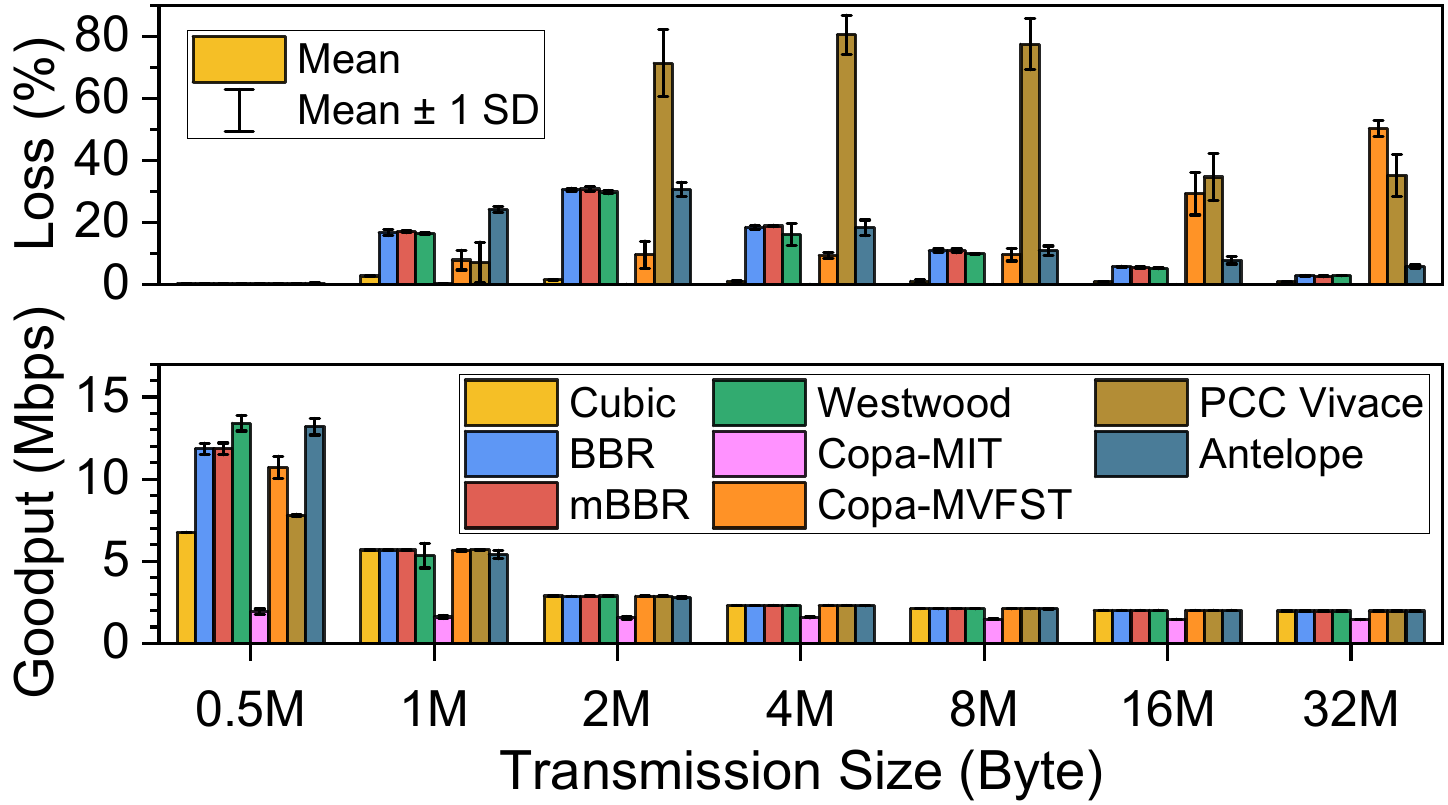}
    \caption{[\rattan{} Emulation] Goodput and packet loss rates of eight CCAs on a mobile network with traffic shaping.}\label{fig:rattan-case-study-more-ccas}
\end{figure}

\section{Case Study}\label{sec:rattan-case-study}
We present a real-world use case illustrating how \rattan{} enables efficient and accurate evaluation of service performance under emerging network conditions that existing tools cannot reproduce.

Mobile Internet Service Providers (ISPs) may deploy traffic policing rules to enforce a maximum allowed speed for users who have exceeded their monthly data plan limits. However, these explicit rate limiters can interfere with the regular operation of Congestion Control Algorithms (CCAs) and lead to unexpected performance degradation.

The authors of mBBR \cite{DBLP:conf/icnp/ZhuLGNL23} analyzed BBR \cite{DBLP:journals/queue/CardwellCGYJ16} performance on real mobile networks and found that BBR is sensitive to traffic policing, resulting in a significantly increased packet loss rate. They designed mBBR for better tolerance of in-network rate limiters and validated its effectiveness on real paths, as shown in \autoref{fig:rattan-case-study-a}. 

However, the mBBR authors found it challenging to investigate how different rate limiter configurations (e.g., target rate, degree of allowed burstiness) impact mBBR's effectiveness. Modifying ISP configurations is impossible, and no existing emulator supports the required traffic policing functionality. As a compromise, they connected a hardware switch providing traffic policing to a link emulator (DummyNet \cite{DBLP:journals/ccr/Rizzo97}) that added a constant delay to create their test environment. This setup is imprecise because the link delay on mobile networks is known to be highly unstable.

Nevertheless, emulating the path with \rattan{} is straightforward: a new cell implementing traffic policing can be inserted into a cellular channel built from existing cells. In our experience, it took a junior intern less than one week to learn the \rattan{} interface and the traffic policing algorithm (i.e., token bucket) to implement this channel. Moreover, writing the cell requires only around 500 lines of code (about 350 of which are reusable boilerplate), and the channel configuration file needs fewer than 50 lines.

We reproduce mBBR's experiments using both their emulation approach (token bucket + constant delay) and \rattan{}'s (token bucket + variable delay). As shown in \autoref{fig:rattan-case-study}, \rattan{}'s results (\autoref{fig:rattan-case-study-c}) closely reproduce the original findings from a real network (\autoref{fig:rattan-case-study-a}). In contrast, the mBBR authors' emulator (\autoref{fig:rattan-case-study-b}) significantly overestimates Cubic's loss rate when the transmission size is small. Further investigation reveals that variable RTTs force Linux's Cubic to exit its slow-start phase early upon detecting increasing RTTs (i.e., the HyStart \cite{DBLP:journals/cn/HaR11} algorithm). Consequently, TCP does not excessively expand its congestion window, resulting in fewer in-flight packets and thus a lower loss rate. When the transmission size is large, there is more time for the congestion window to converge to a stable value, resulting in a lower and more accurate mean loss rate that is closer to that of the real network. This result shows that \rattan{}'s flexibility allows for more fine-grained modeling of network conditions and, consequently, more accurate emulation.

Using \rattan{}, we can easily investigate how traffic policing affects a wider range of CCAs. As shown in \autoref{fig:rattan-case-study-more-ccas}, we test five additional CCAs from different categories: Westwood \cite{DBLP:conf/mobicom/MascoloCGSW01} (based on bandwidth estimation), Copa \cite{DBLP:conf/nsdi/ArunB18} (based on latency, with two different configurations: Copa-MIT \cite{venkatarun_ccp_copa_2018} and Copa-MVFST \cite{facebook_mvfst_2020}), PCC Vivace \cite{DBLP:conf/nsdi/DongMZAGGS18} (based on online learning), and Antelope \cite{DBLP:conf/icnp/ZhouQLTLDW21} (reinforcement learning driven CCA selection). The results clearly show that different CCAs (or the same CCA with varied parameters) exhibit markedly different sensitivities to the rate limiter. Copa-MIT is unable to fully utilize the available 2~Mbps bandwidth because it overly reduces its rate, tries to meet its strict delay control target under high link latency fluctuation. Copa-MVFST, in contrast, experiences an extremely high packet loss rate because it allows a much larger number of in-flight packets, which can exceed traffic policing's allowance of burst. Antelope and Westwood both achieve good performance, comparable to that of mBBR, and exhibit good compatibility with traffic policing. PCC Vivace, however, shows a high loss rate with medium transmission sizes because its probing-deciding process takes time to converge and may overshoot during its exploration.

Beyond different CCAs, we can also use \rattan{} to investigate hypothetical network path scenarios. For example, WAN optimizers often apply traffic shaping\footnote{While both throttle throughput, traffic policing drops all excessive packets immediately, whereas traffic shaping buffers them.} for their QoS policies~\cite{DBLP:conf/sigcomm/LiNCGM19}. We explore how traffic shaping interferes with CCAs using \rattan{}. As the results in \autoref{fig:rattan-case-study-different-tokenbucket} show, mBBR is not effective in the traffic-shaping case. The loss rate of Copa-MVFST is lower for small transmission sizes, as traffic shaping allows more packets to be buffered. All other CCAs achieve performance similar to that in the traffic-policing case. This result highlights \rattan{}'s potential to validate the generality of transport innovations like mBBR.

Notably, the experiments for \autoref{fig:rattan-case-study-different-tokenbucket} and \autoref{fig:rattan-case-study-more-ccas} were designed, implemented, and conducted in a single day with the help of the \rattan{} orchestrator, showcasing the flexibility and user-friendliness of \rattan{}.

\section{Conclusion}\label{sec:conclusion}
The Internet's increasing complexity and dynamics pose significant challenges to existing network emulators in terms of flexibility, scalability, and usability. To address this, we present \rattan{}, a high-performance, modular framework for large-scale Internet path emulation that supports complex emulation models. By empowering developers with a robust and extensible evaluation platform, we believe \rattan{} will significantly accelerate the development of next-generation network innovations.

\clearpage
\bibliographystyle{ACM-Reference-Format}
\bibliography{ref.bib}

\end{document}